\documentclass[journal=jacsat,manuscript=article]{achemso}

\usepackage[version=3]{mhchem} 
\usepackage{amsmath, amssymb}
\usepackage{mathrsfs}
\usepackage[utf8]{inputenc}

\author{Eirik F.~Kjønstad}
\affiliation{Department of Chemistry, Norwegian University of Science and Technology, 7491 Trondheim, Norway} 
\author{Henrik Koch}
\affiliation{Scuola normale superiore, Piazza dei Cavalieri, 7, 56126 Pisa PI, Italy}
\alsoaffiliation{Department of Chemistry, Norwegian University of Science and Technology, 7491 Trondheim, Norway} 
\email{henrik.koch@sns.it}

\title{A biorthonormal formalism for nonadiabatic coupled cluster dynamics}

\abbreviations{IR,NMR,UV}

\newcommand{\bsym}{\boldsymbol} 

\renewcommand{\b}{\bsym}
\newcommand{\adj}{^\dagger}

\newcommand{\ket}[1]{\vert #1 \rangle}

\newcommand{\bra}[1]{\langle #1 \vert}

\newcommand{\braket}[1]{\langle #1 \rangle}

\newcommand{\Dbraket}[2]{\langle #1 \hspace{.10em} \vert \hspace{.10em}  #2 \rangle}
\newcommand{\Tbraket}[3]{\langle #1 \hspace{.10em} \vert \hspace{.10em} #2 \hspace{.10em} \vert \hspace{.10em} #3 \rangle}

\newcommand{\DBraket}[2]{\Big\langle #1 \hspace{.10em} \Big\vert \hspace{.10em}  #2 \Big\rangle}

\newcommand{\hf}{\mathrm{HF}}
\newcommand{\cc}{\mathrm{CC}}

\newcommand{\st}[1]{\mathcal{#1}}
\newcommand{\bst}[1]{\b{\mathcal{#1}}}

\newcommand{\df}{\mathrm{d}}

\begin{document}

\begin{abstract}
    In coupled cluster methods, the electronic states are biorthonormal in the sense that the left states are orthonormal to the right states. Here we present an extension of this formalism to a left and right  total molecular wave function. Starting from left and right Born-Huang expansions, we derive projected Schrödinger equations for the left and right nuclear wave functions. Observables may be extracted from the resulting wave function pair using standard expressions. The formalism is shown to be invariant under electronic basis transformations, such as normalization of the electronic states. Consequently, the nonadiabatic coupling elements can be expressed with biorthonormal wave functions. Calculating normalization factors that scale as full-CI is therefore not necessary, contrary to claims in the literature. For nuclear dynamics, we therefore need expressions for the vector and scalar couplings in the biorthonormal formalism. We derive these expressions using a Lagrangian formalism.
\end{abstract}

\section{Introduction}

Nonadiabatic coupling elements account for electron-nucleus interactions that are neglected in the Born-Oppenheimer\cite{Born1927} (BO) approximation. These elements couple different electronic states through the nuclear kinetic energy operator. While mostly negligible in ground state chemistry, coupling elements are required when considering molecular dynamics in excited electronic states. Excited state dynamics often involves regions of nuclear space where  electronic states are nearly or exactly degenerate, causing a  breakdown of the BO separation.\cite{Zhu2016,Curchod2018} Accurately describing nonadiabatic coupling elements is therefore important for reliable predictions in photochemistry.

The coupled cluster method is one of the most accurate electronic structure methods, both for ground and excited state properties,\cite{bartlett2007coupled,Krylov2008,helgaker2014molecular,Loos2020a} but it has not found widespread use for predicting excited state dynamics. This is primarily because standard coupled cluster methods give a nonphysical description of regions close to electronic degeneracies, or conical intersections\cite{hattig2005structure,kohn2007can,kjoenstad2017crossing}. This issue can be traced to the method's non-Hermiticity, which seems to imply that coupled cluster methods cannot be used for nonadiabatic dynamics. However, this is not the case. As we have shown in recent work, the method can be constrained to give a correct physical description of excited state conical intersections while retaining the standard non-Hermitian formalism and presumably its accuracy\cite{kjoenstad2017resolving,kjoenstad2019invariant}. These developments may lead to renewed interest in coupled cluster dynamics.

Nonadiabaticity, as described by coupled cluster methods, has been considered by several authors. The formula for the vector coupling was first derived by Christiansen,\cite{Christiansen1999nonadiabatic} who applied the $Z$-vector substitution method\cite{handy1984evaluation} on a biorthonormal expression for the vector coupling,
\begin{align}
    \b{F}_{mn}^I = \Dbraket{\tilde{\psi}_m}{\nabla_I \psi_n}, \quad \Dbraket{\tilde{\psi}_m}{\psi_n} = \delta_{mn}, \label{eq:biorthonormal_F}
\end{align}
where $(\tilde{\psi}_k,\psi_k)$ refers to the left and right $k$th electronic states, and $I$ to a nucleus.
However, Christiansen's paper\cite{Christiansen1999nonadiabatic} did not include an implementation of the coupling. The vector coupling was later rederived by Tajti and Szalay\cite{Tajti2009} by differentiating the corresponding  $m$-to-$n$ transition element of the electronic Hamiltonian. Their derivation is closely related to that given by Ichino \emph{et al.}\cite{Ichino2009} for the quasidiabatic interstate coupling. Tajti and Szalay\cite{Tajti2009} also gave an implementation of the vector coupling at the singles and doubles level (CCSD\cite{Purvis1982}). These papers on the vector coupling\cite{Christiansen1999nonadiabatic,Tajti2009} did not include a discussion of the nuclear Schrödinger equations in coupled cluster theory, where the coupling elements enter.

The correct formula for the vector coupling has been a subject of some controversy. 
Tajti and Szalay\cite{Tajti2009} argued that the biorthonormal formula in Eq.~\eqref{eq:biorthonormal_F} is incorrect. As they correctly noted, the vector coupling changes with the norm of the left and right states. A similar observation had been made in an earlier paper on the diagonal BO correction.\cite{Gauss2006}  
Since the vector coupling varies with the norm of the states, the full-CC vector coupling is different from the full configuration interaction (CI) limit, where left and right states are identical and usually  normalized. They therefore suggested that normalizing the states was necessary. Furthermore, since the derivative can either act on the left or on the right state, they suggested using an average of the two\cite{Tajti2009}. If true, these observations are troubling: they imply that computing the vector coupling has a computational cost that scales as full-CI due to the normalization factors for the right states. In practice, the normalization factors are therefore approximated. However, it is unfortunate if one must resort to approximations other than the truncation level of the coupled cluster method (e.g., singles and doubles). The need for normalization factors was also assumed in the recent CCSD implementation by Faraji \emph{et al}.\cite{Faraji2018} 

One of the main objectives of the present paper is to establish that normalization is not necessary. The reason is that normalization is a special case of an invertible transformation of the electronic basis. Such transformations do not change the expansion space in the Born-Huang expansion\cite{Born1954} 
and therefore do not change the molecular wave function. In particular, the coefficients in the Born-Huang expansion---that is, the nuclear wave functions---absorb the transformation of the electronic states. In a recent paper, Shamasundar\cite{Shamasundar2018} also noted that the predicted dynamics must not depend on the normalization of the underlying electronic wave functions. The vector coupling does depend on normalization, but this should not be considered a problem because this quantity is not an observable. Since normalization of wave functions is not necessary, the biorthonormal formula in Eq.~\eqref{eq:biorthonormal_F} is a valid option. In this work, we derive the biorthonormal coupled cluster vector and scalar couplings using the Lagrangian approach developed by Hohenstein.\cite{Hohenstein2016}

The second main objective of the paper is to give a framework for nonadiabatic dynamics using coupled cluster methods. In particular, we argue that the biorthonormal formalism for electronic wave functions implies a biorthonormal formalism for the molecular wave function. Hence, we must determine left and right nuclear wave functions and the nuclear motion is described by two sets of nuclear Schrödinger equations. The result is a molecular wave function pair ($\tilde{\Psi}, \Psi$), where observable quantities are given by the usual biorthonormal formulas. In this contribution, we describe theoretical aspects relevant for nonadiabatic dynamics. Implementation of the various quantities is postponed to a future publication.

\section{Theory}
The total wave function of a molecular system can be expressed as an expansion over the electronic wave functions. The coefficients of this Born-Huang expansion defines the nuclear wave functions. These are determined by inserting the expansion in the Schrödinger equation and projecting out the electronic components. To formulate the corresponding procedure for coupled cluster theory, we first review the description of the electronic states.

\subsection{Electronic wave functions in coupled cluster theory}
In the equation of motion coupled cluster formalism,
a set of left and right electronic states are considered. These are defined as\citep{Stanton1993}
\begin{align}
    \bra{\tilde{\psi}_n} &= \sum_{\mu \geq 0} \st{L}_{\mu}^n \bra{\mu} \exp(-T), \label{eq:EOM_left} \\
    \ket{\psi_n} &= \sum_{\mu \geq 0} \exp(T) \ket{\mu} \st{R}_\mu^n, \quad n = 0, 1, 2, \ldots \label{eq:EOM_right} 
\end{align}
The states are not identical, in general, but they satisfy the biorthonormality condition
\begin{align}
    \Dbraket{\tilde{\psi}_m}{\psi_n} = \delta_{mn}. \label{eq:state_biorthonormality}
\end{align}
The scalars $\st{L}_\mu^n$ and $\st{R}_\mu^n$ are state amplitudes, giving weights to the configurations 
\begin{align}
    \ket{\mu} &= \tau_\mu \ket{\hf} \\
    \bra{\mu} &= \bra{\hf} \tilde{\tau}_{\mu}^\dagger, \quad \mu \geq 0,
\end{align}
where $\tau_\mu$ and $\tilde{\tau}_{\mu}$ with $\mu > 0$ are excitation operators relative to the Hartree-Fock state $\ket{\hf}$, while $\tau_0 = \tilde{\tau}_0 = \mathbb{I}$ is the identity operator. The ket and bra bases, $\{ \ket{\mu} \}$ and $\{ \bra{\mu}\}$, span the same subspace and are normally required to satisfy the biorthonormality relation
\begin{align}
\Dbraket{\mu}{\nu} = \delta_{\mu\nu}, \quad \mu,\nu \geq 0.
\end{align}
One special case is that the left and right bases are identical and hence orthonormal. However, using different left and right basis is sometimes convenient (e.g., in spin-adapted formulations\citep{helgaker2014molecular}).
Finally, we have the exponential part of the parametrization, defined by the cluster operator
\begin{align}
    T = \sum_{\mu > 0} t_\mu \tau_\mu.
\end{align}
The scalars $t_\mu$ are called cluster amplitudes. 

Given the parametrization in Eqs.~\eqref{eq:EOM_left} and \eqref{eq:EOM_right}, how are the parameters determined? First one assumes that the right ground state can be written
\begin{align}
    \ket{\psi_0} = \exp(T) \ket{\hf}.
\end{align}
Then the time-independent Scrödinger equation, expressed as
\begin{align}
    \exp(-T) H \exp(T)\ket{\hf} = E_0 \ket{\hf},
\end{align}
is projected onto the bra basis. The operator
\begin{align}
    \bar{H} = \exp(-T) H \exp(T)
\end{align}
is known as the similarity transformed Hamiltonian. This projection procedure gives an expression for the ground state energy and equations for determining the amplitudes,
\begin{align}
    E_0 &= \Tbraket{\hf}{\bar{H}}{\hf} \label{eq:E_0} \\
    \Omega_\mu  &= \Tbraket{\mu}{\bar{H}}{\hf} = 0, \quad \mu > 0.
\end{align}
The state amplitudes are determined by making the pseudo expectation values
\begin{align}
    E_n(\bst{L}_n, \bst{R}_n) = \Tbraket{\tilde{\psi_n}}{\bar{H}}{\psi_n}, \quad n = 0,1,2,\ldots \label{eq:pseudo_exp}
\end{align}
stationary under the binormality condition given in Eq.~\eqref{eq:state_biorthonormality}. This constrained optimization problem is conveniently formulated in terms of the state Lagrangians
\begin{align}
\begin{split} 
    \mathscr{L}_n(\bst{L}_n, \bst{R}_n, \bar{E}_n) &= \Tbraket{\tilde{\psi_n}}{\bar{H}}{\psi_n} + \bar{E}_n (1- \Dbraket{\tilde{\psi}_n}{\psi_n}) \\
    &= \bst{L}_n^T \bar{\b{\mathcal{H}}} \bst{R}_n + \bar{E}_n (1 - \bst{L}_n^T \bst{R}_n),
\end{split}
\end{align}
where we have defined the Hamiltonian matrix
\begin{align}
    \bar{\mathcal{H}}_{\mu\nu} = \Tbraket{\mu}{\bar{H}}{\nu}, \quad \mu,\nu \geq 0.
\end{align}
Stationarity of the Lagrangians imply
\begin{align}
    1 &= \bst{L}_n^T \bst{R}_n \\
    \b{0} &= \bst{L}_n^T \bar{\b{\mathcal{H}}} - \bar{E}_n \bst{L}_n^T \\ 
    \b{0} &= \bar{\b{\mathcal{H}}} \bst{R}_n - \bar{E}_n \bst{R}_n.
\end{align}
Writing $\bar{E}_n = E_n = E_0 + \omega_n$, we see that the latter two equations read
\begin{align}
    \bst{L}_n^T \bst{A} &= \omega_n \bst{L}_n^T \\
    \bst{A} \bst{R}_n   &= \omega_n \bst{R}_n, \quad n = 0,1,2,\ldots,
\end{align}
where $\bst{A} = \bar{\bst{H}} - E_0 \b{I}$. Identifying the pseudo expectation value in Eq.~\eqref{eq:pseudo_exp} with the energy, we see that $\omega_n$ is the excitation energy of the $n$th state, where it is understood that $\omega_0 = 0$ for the ground state. The matrix $\bst{A}$ can be expressed as
\begin{align}
    \bst{A} = \begin{pmatrix} 
                0     & \b{\eta}^T \\
                \b{0} & \b{A}
            \end{pmatrix}
\end{align}
where
\begin{align}
    \eta_\nu   &= \Tbraket{\hf}{[\bar{H}, \tau_\nu]}{\hf} \\
    A_{\mu\nu} &= \Tbraket{\tilde{\mu}}{[\bar{H}, \tau_\nu] }{\hf}.
\end{align}
where $\b{A}$ is called the coupled cluster Jacobian matrix.\citep{Koch1990} The eigenvalues of $\b{A}$ are the non-zero excitation energies, i.e.~$\omega_n$ with $n = 1,2,3,\ldots$

\subsection{The Born-Huang expansion of the total wave function and the nuclear Schrödinger equations}
With the electronic states described, we now turn to the expansion of the total wave function.
The Born-Huang expansion expresses the total wave function in terms of the left and right electronic bases given in Eqs.~\eqref{eq:EOM_left} and \eqref{eq:EOM_right}. Notice that this implies a biorthonormal description of the total wave function, since we can expand in both the left and right states. Hence we have a left and a right total wave function
\begin{align}
    \Psi(\b{r}, \b{R}, t) &= \sum_n \chi_n(\b{R},t) \psi_n(\b{r}; \b{R}) \label{eq:state_right} \\
    \tilde{\Psi}(\b{r}, \b{R}, t) &= \sum_n \tilde{\chi}_n(\b{R},t) \tilde{\psi}_n(\b{r}; \b{R}), \label{eq:state_left}
\end{align}
with associated left and right nuclear wave functions $\tilde{\chi}_n$ and $\chi_n$, and
\begin{align}
    1 = \Dbraket{\tilde{\Psi}}{\Psi} = \sum_{mn} \Tbraket{\tilde{\chi}_m}{\Dbraket{\tilde{\psi_m}}{\psi_n}}{\chi_n} = \sum_{mn} \delta_{mn} \Dbraket{\tilde{\chi}_m}{\chi_n} = \sum_n \Dbraket{\tilde{\chi}_n}{\chi_n},
\end{align}
where we have assumed biorthonormal electronic states in the third equality.
Expectation values are defined through the standard expression\citep{Koch1990,Stanton1993}
\begin{align}
    \braket{\Omega} = \Tbraket{\tilde{\Psi}}{\Omega}{\Psi}, \quad \Omega = \Omega\adj. \label{eq:expectation_value}
\end{align}

To derive the equations for the nuclear wave functions, one normally projects the total Schrödinger equation on the electronic basis. In this respect, a biorthonormal description is advantageous; for practical coupled cluster models, where the excitation space is truncated to some excitation order, projection of the \emph{right} Schrödinger equation is done onto the \emph{left} electronic basis, leading to computationally tractable expressions that scale as expected for the given model (e.g.~$\mathcal{O}(N^6)$ for CCSD). 

By inserting the $\Psi$ in Eq.~\eqref{eq:state_right} into the time dependent Schrödinger equation,
\begin{align}
    H \Psi = i \frac{\df \Psi}{\df t},
\end{align}
and projecting it onto the left electronic basis, we get a coupled set of equations for the right nuclear wave functions $\chi_n$. 
These nuclear Schrödinger equations can be expressed as
\begin{align}
    (i \frac{\df}{\df t} - E_m) \chi_m = \sum_{I,n} \frac{1}{2 M_I} (\delta_{mn}\nabla_I^2 + G_{mn}^I + 2 \b{F}_{mn}^I \cdot \nabla_I) \chi_{n}, \label{eq:nuclear_SE_right}
\end{align}
where we have suppressed the $\b{R}$ and $t$ dependence for readability. The nonadiabatic coupling vectors in Eq.~\eqref{eq:nuclear_SE_right} are given in the biorthonormal basis:
\begin{align}
    G_{mn}^I &= \Dbraket{\tilde{\psi}_m}{\nabla_I^2 \psi_n} \label{eq:scalar_coupling} \\
    \b{F}_{mn}^I &= \Dbraket{\tilde{\psi}_m}{\nabla_I \psi_n}. \label{eq:vector_coupling}
\end{align}
These are called the scalar and vector couplings, respectively.

In analogous fashion, we derive the nuclear Schrödinger equations for the left nuclear wave functions from the complex conjugated Schrödinger equation
\begin{align}
    H \tilde{\Psi} = -i \frac{\df \tilde{\Psi}}{\df t}. \label{eq:conjugated_TDSE}
\end{align}
Inserting Eq.~\eqref{eq:state_left} into Eq.~\eqref{eq:conjugated_TDSE}, and projecting onto the left electronic basis, leads to
\begin{align}
    (-i \frac{\df}{\df t} - E_m) \tilde{\chi}_m = \sum_{I,n} \frac{1}{2 M_I} (\delta_{nm}\nabla_I^2 + \tilde{G}_{nm}^I + 2 \tilde{\b{F}}_{nm}^I \cdot \nabla_I) \tilde{\chi}_{n}, \label{eq:nuclear_SE_left}
\end{align}
where
\begin{align}
    \tilde{G}_{nm}^I &= \Dbraket{\nabla_I^2 \tilde{\psi}_n}{\psi_m} \\
    \tilde{\b{F}}_{nm}^I &= \Dbraket{\nabla_I \tilde{\psi}_n}{\psi_m}.
\end{align}

The nuclear Schrödinger equations may be expressed in the more compact matrix notation
\begin{align}
    (i \frac{\df }{\df t} - \b{E}) \b{\chi} &= \sum_I \frac{1}{2 M_I} (\b{I}\nabla_I^2 + \b{G}_I + 2 \b{F}_I \cdot \nabla_I)\b{\chi} \label{eq:right_nuclear_SE_matrix} \\
    (-i \frac{\df }{\df t} - \b{E}) \tilde{\b{\chi}}  &= \sum_I \frac{1}{2 M_I} (\b{I}\nabla_I^2 + \tilde{\b{G}}_I + 2 \tilde{\b{F}}_I \cdot \nabla_I)\tilde{\b{\chi}},
\end{align}
where $\b{E}$ is a diagonal matrix with the electronic energies on the diagonal, $\b{I}$ is the identity matrix, $\b{\chi}$ is a vector containing the right nuclear wave functions, and $\b{G}_I$ and $\b{F}_I$ are matrices consisting of the scalar and vector couplings of the $I$th nucleus, respectively. The quantities with a tilde are similarly defined. 

This matrix notation has been used to illuminate some relations to gauge theories in the nuclear Schrödinger equations; Pacher \emph{et al}.\citep{Pacher1989} found that the vector coupling can be seen to serve a role analogous to the vector potential in electromagnetism. In the present work, it serves as a useful notation for dealing with basis transformations and the vector algebra needed to demonstrate invariance under such transformations.

\subsection{Basis invariance and the special case of norm invariance}
In the literature on nonadiabatic coupling vectors in coupled cluster theory, normalization is often considered problematic. The reason is that the left and right states are binormal in the coupled cluster formalism. Compared to the nonadiabatic couplings in full-CI theory, where the states are normalized, the full coupled cluster limit is ``incorrect'' because the value of the couplings depend on the geometry-dependent normalization constants. While this suggests that one should normalize the states, doing so is not straightforward. The computational cost of the normalization factor scales as full-CI for the right electronic states:\citep{Ichino2009}
\begin{align}
    N_R^n &= \Dbraket{\psi_n}{\psi_n} \\
    N_L^n &= \Dbraket{\tilde{\psi}_n}{\tilde{\psi}_n}.
\end{align}
Since one cannot evaluate $N_R^n$ in general, some have suggested $N_R^n = (N_L^n)^{-1}$ or $N_R^n = N_L^n = 1$ as alternatives. The former gives the full-CI limit while the latter simply assumes the standard binormality.\citep{Ichino2009,Faraji2018}

Binormality is not an issue from the point of view of dynamics. Changing the norm of the electronic states is a special case of a basis transformation of the electronic basis. As such, the Born-Huang expansion and the projection equations are equivalent in the transformed and untransformed bases. Changes in the electronic basis are absorbed in the expansion coefficients, i.e., the nuclear wave functions. In the special case of normalization, the right electronic wave functions are divided by $N_R^n$ while the right nuclear wave functions are multiplied by $N_R^n$. The total wave function is invariant under such transformations.

More precisely, consider invertible transformations of the left and right electronic bases. In vector notation, these transformations can be expressed as
\begin{align}
    \tilde{\b{\psi}}' &= \tilde{\b{\psi}} \b{N} \\
    \b{\psi}' &= \b{\psi} \b{M},
\end{align}
where the matrices $\b{M}$ and $\b{N}$ are assumed to be smooth invertible matrix functions of the nuclear coordinates. For notational simplicity, we have let the left and right wave function vectors be row vectors. Transformed quantities are denoted by a prime.
In the transformed basis, the total left and right wave function have the Born-Huang expansions
\begin{align}
        \Psi'(\b{r}, \b{R}, t) &= \sum_n \chi_n'(\b{R},t) \psi_n'(\b{r}; \b{R}) \label{eq:state_right_transformed} \\ 
        \tilde{\Psi}'(\b{r}, \b{R}, t) &= \sum_n \tilde{\chi}_n'(\b{R},t) \tilde{\psi}_n'(\b{r}; \b{R}). \label{eq:state_left_transformed}
\end{align}
We wish to show that the wave function in the transformed basis is identical to that obtained in the untransformed basis; that is, $\Psi' = \Psi$ and $\tilde{\Psi}' = \tilde{\Psi}$. The conclusion that follows is that the choice of electronic basis does not change the predictions of the theory. In other words, it is perfectly appropriate to use the biorthonormal description that is standard in coupled cluster theory.\citep{Koch1990}

Before proceeding, we define some notation. In the transformed basis, 
we have to account for the non-unit overlap of the electronic wave functions. Hence, when projecting the time-dependent Schrödinger equation onto the electronic basis, we get electronic overlap matrix elements. We define these elements as
\begin{align}
    S_{mn} = \Dbraket{\tilde{\psi}_m'}{\psi_n'} = \sum_{kl} \Dbraket{\tilde{\psi}_k N_{km}}{\psi_l M_{ln}} = \sum_{kl} N_{km}^\ast \delta_{kl} M_{ln} = (\b{N}\adj \b{M})_{mn}.
\end{align}
Similarly, the electronic Hamiltonian matrix is not necessarily diagonal:
\begin{align}
    (\b{H}_e)_{mn} = \Dbraket{\tilde{\psi}_m'}{H_e \psi_n'} = \sum_{kl} \Dbraket{\tilde{\psi}_k N_{km}}{H_e \psi_l M_{ln}} = \sum_{kl} N_{km}^\ast E_{kl} M_{ln} = (\b{N}\adj \b{E} \b{M})_{mn}.
\end{align}

We show the equivalence for the right wave functions. The proof for the left wave function is identical. Following the standard procedure, we now insert the transformed wave function in Eq.~\eqref{eq:state_right_transformed} into the Schrödinger equation and project onto the transformed left electronic wave functions.  
The result is the right nuclear Schrödinger equation
\begin{align}
    \bigl( \b{N}\adj \b{E} \b{M} - i \frac{\df}{\df t} \b{S}  \bigr)\b{\chi}' = \sum_I \frac{1}{2 M_I} (\b{S} \nabla_I^2  + \b{G}_I' + 2 \b{F}_I' \cdot \nabla_I) \b{\chi}'. \label{eq:right_nuclear_SE_transformed}
\end{align}
If the total wave function is invariant, and $\b{\psi}' = \b{\psi} \b{M}$, then we must have nuclear wave functions that cancel the transformation of the electronic wave functions:
\begin{align}
    \b{\chi}' = \b{M}^{-1} \b{\chi}. \label{eq:transformed_nuclear_wave_functions}
\end{align}
Indeed, with $\b{\chi}'$ as given in Eq.~\eqref{eq:transformed_nuclear_wave_functions}, we have
\begin{align}
    \Psi' = \sum_k \psi_k' \chi_k' = \sum_{klm} \psi_l M_{lk} M_{km}^{-1} \chi_m = \sum_l \psi_l \chi_l = \Psi.
\end{align}

Let us confirm that Eq.~\eqref{eq:transformed_nuclear_wave_functions} is in fact a solution to the transformed nuclear Schrödinger equation in Eq.~\eqref{eq:right_nuclear_SE_transformed}. We begin by relating the old and new nonadiabatic coupling terms. The gradient of the electronic wave functions transform as
\begin{align}
    \nabla_I \psi_l' = \sum_m \nabla_I (\psi_m M_{ml}) = \sum_m \Bigl( (\nabla_I \psi_m) M_{ml} + \psi_m (\nabla_I M_{ml}) \Bigr).
\end{align}
Hence, the vector couplings can be written as
\begin{align}
    (\b{F}_I')_{kl} = \Dbraket{\tilde{\psi}_k'}{\nabla_I \psi_l'} = \sum_n N_{nk}^\ast \Dbraket{\tilde{\psi}_n}{\nabla_I \psi_l'} = \sum_{nm} \Bigl( N_{nk}^\ast F_{nm}^I M_{ml} + N_{nk}^\ast \delta_{nm} \nabla_I M_{ml} \Bigr).
\end{align}
In more compact matrix notation, we have
\begin{align}
    \b{F}_I' = \b{N}\adj \b{F}_I \b{M} + \b{N}\adj (\nabla_I \b{M}).
\end{align}
Similarly, the Laplacian of the electronic wave functions transform as
\begin{align}
    \nabla_I^2 \psi_l' = \nabla_I \cdot \nabla_I \psi_l' = \sum_m \Bigl( (\nabla_I^2 \psi_m) M_{ml} + 2 (\nabla_I \psi_m)\cdot (\nabla_I M_{ml}) + \psi_m (\nabla_I^2 M_{ml}) \Bigr),
\end{align}
implying that the scalar couplings transform as
\begin{align}
    \b{G}_I' = \b{N}\adj \b{G}_I \b{M} + 2 \b{N}\adj \b{F}_I \cdot (\nabla_I \b{M}) + \b{N}\adj (\nabla_I^2 \b{M}).
\end{align}

The gradient and Laplacian of $\b{\chi}'$ is derived in the same way as for the electronic states, giving
\begin{align}
    \nabla_I \chi_l' &= \sum_m \bigl( M_{lm}^{-1} (\nabla_I \chi_m) +  (\nabla_I M_{lm}^{-1})\chi_m \bigr) \\
    \nabla_I^2 \chi_l' &= \sum_m \bigl( M_{lm}^{-1} (\nabla_I^2 \chi_m) + 2  (\nabla_I M_{lm}^{-1}) \cdot (\nabla_I \chi_m) + (\nabla_I^2 M_{lm}^{-1})\chi_m \bigr).
\end{align}
Thus, we have the following contributions on the right hand side of the nuclear Schrödinger equation:
\begin{align}
    \b{S} \nabla_I^2 \b{\chi}' &= \b{N}\adj (\nabla_I^2 \b{\chi} + 2 \b{M} (\nabla_I \b{M}^{-1}) \cdot \nabla_I \b{\chi} + \b{M} (\nabla_I^2 \b{M}^{-1}) \b{\chi} ) \\
    \b{G}_I' \b{\chi}' &= \b{N}\adj (\b{G}_I \b{\chi} + 2 \b{F}_I \cdot (\nabla_I \b{M})\b{M}^{-1}\b{\chi} + (\nabla_I^2 \b{M})\b{M}^{-1}\b{\chi}) \\
\begin{split}
    2 \b{F}_I'\cdot\nabla_I\b{\chi}' &= \b{N}\adj ( 2 \b{F}_I \cdot \nabla_I \b{\chi} + 2 \b{F}_I \cdot \b{M} (\nabla_I \b{M}^{-1}) \b{\chi} \\
    &\quad\quad 2 (\nabla_I \b{M})\b{M}^{-1} \cdot \nabla_I \b{\chi} + 2 (\nabla_I\b{M})\cdot (\nabla_I \b{M}^{-1})\b{\chi}).
\end{split}
\end{align}
Though somewhat involved, most of the terms cancel when added together. In fact, since
\begin{align}
    0 &= \nabla_I (\b{M} \b{M}^{-1}) = (\nabla_I \b{M}) \b{M}^{-1} + \b{M} (\nabla_I \b{M}^{-1}) \\ 
    0 &= \nabla_I^2 (\b{M} \b{M}^{-1}) = (\nabla_I^2 \b{M}) \b{M}^{-1} + \b{M} (\nabla_I^2 \b{M}^{-1}) + 2 (\nabla_I \b{M}) \cdot (\nabla_I \b{M}^{-1}),
\end{align}
we can write
\begin{align}
\begin{split}
    \b{S} \nabla_I^2 \b{\chi}' + \b{G}_I' \b{\chi}' + 2 \b{F}_I'\cdot\nabla_I\b{\chi}' &= \b{N}\adj \bigl( 
                                     \nabla_I^2 \b{\chi} + \b{G}_I \b{\chi} + 2 \b{F}_I \cdot \nabla_I \b{\chi} \\
                                &+ 2 \nabla_I ( \b{M} \b{M}^{-1} ) \cdot \nabla_I \b{\chi} + \nabla_I^2 ( \b{M} \b{M}^{-1} )\b{\chi} \\
                                   &+  2 \b{F}_I \cdot \nabla_I ( \b{M} \b{M}^{-1} ) \b{\chi} \bigr) \\
                                   &= \b{N}\adj ( 
                                     \nabla_I^2 \b{\chi} + \b{G}_I \b{\chi} + 2 \b{F}_I \cdot \nabla_I \b{\chi}).
\end{split}
\end{align}
In other words, with $\b{\chi}' = \b{M}^{-1} \b{\chi}$, the right nuclear Schrödinger equation simplifies to
\begin{align}
    \bigl( \b{N}\adj \b{E} - i \frac{\df}{\df t} \b{N}\adj  \bigr)\b{\chi} = \b{N}\adj \sum_I \frac{1}{2 M_I} ( \nabla_I^2  + \b{G}_I + 2 \b{F}_I \cdot \nabla_I) \b{\chi}, \label{eq:right_nuclear_SE_transformed_2}
\end{align}
which, upon premultiplication by $\b{N}^{-\dagger}$, is seen to be equivalent to the original right nuclear Schrödinger equation in Eq.~\eqref{eq:right_nuclear_SE_matrix}. 

Since all the derivation steps we have made are reversible, we have shown that $\b{\chi}$ is a solution to the untransformed nuclear Schrödinger equation if and only if $\b{\chi}'$ is a solution to the transformed Schrödinger equation. The total right wave function is therefore invariant with respect to transformations of the electronic basis, $\Psi' = \Psi$. 

One consequence of basis invariance is that the nonadiabatic couplings can be derived in the standard biorthonormal formalism. To  derive expressions for these elements, we must first consider the geometry dependence of the many-body  operators.

\subsection{Geometry dependence of the many-body operators}
The scalar and vectors couplings, see Eqs.~\eqref{eq:scalar_coupling} and \eqref{eq:vector_coupling}, involve differentiation of the electronic wave functions with respect to the nuclear coordinates $\b{x}$. To evaluate these, we need to consider the dependence of both the wave function parameters and the many-body operators. The operator's dependence is handled through orbital connections which relates orbitals at neighbouring geometries. Note that there is no unique orbital connection; many-body operators are expressed with respect to a specific orthonormal orbital basis, but at each geometry there are an infinite number of such bases related by unitary transformations. For reasons that will become clear, we will use the so-called natural connection. Our presentation will follow closely that given by of Olsen \emph{et al}.\cite{olsen1995orbital}

When evaluating derivatives at $\b{x}_0$, we need to relate the basis at $\b{x}_0$ to some basis at $\b{x} = \b{x}_0 + \Delta \b{x}$. Suppose the molecular orbitals (MOs) at $\b{x}_0$ are 
\begin{align}
    \phi_m(\b{x}_0) = \sum_\alpha  \chi_\alpha(\b{x}_0) C_{\alpha m}(\b{x}_0),
\end{align}
where $C_{\alpha m}$ are orbital coefficients and  $\chi_\alpha$ are atomic orbitals. The unmodified MOs (UMOs) are defined by freezing the orbital coefficients,
\begin{align}
    \phi_m^u(\b{x}) = \sum_\alpha C_{\alpha m}(\b{x}_0) \chi_\alpha(\b{x}).
\end{align}
The UMOs are not orthonormal, however:
\begin{align}
    S_{mn}(\b{x}) = \Dbraket{\phi_m^u(\b{x})}{\phi_n^u(\b{x})}, \quad S_{mn}(\b{x}_0) = \delta_{pq}.
\end{align}
Hence, UMOs are related to orthonormalized MOs (or OMOs) through
\begin{align}
    \phi_m(\b{x}) = \sum_n \phi_n^u(\b{x}) T_{nm}(\b{x}),
\end{align}
where the connection matrix $\b{T}(\b{x})$ satisfies $\b{T}(\b{x}_0) = \b{I}$ and
\begin{align}
    \b{T}(\b{x})\adj \b{S}(\b{x}) \b{T}(\b{x}) = \b{I}. \label{eq:orthogonality_connection_mat}
\end{align}
In the natural connection, $\b{T}$ is chosen to be
\begin{align}
    \b{T}(\b{x}) = \b{W}(\b{x})^{-1} (\b{W}(\b{x}) \b{S}(\b{x}) \b{W}(\b{x})\adj)^{1/2} = \b{W}(\b{x})^{-1} \b{\Delta}(\b{x}),
\end{align}
where
\begin{align}
    W_{mn}(\b{x}) = \Dbraket{\phi_m^u(\b{x}_0)}{\phi_n^u(\b{x})}.
\end{align}
The natural connection minimizes the change in the orthonormalized orbitals at $\b{x}$ relative to the orbitals at $\b{x}_0$. 

Let us now relate the orbital space at $\b{x}$ to the orbital space at $\b{x}_0$. In order to do so, we need to consider a \emph{complete} orbital basis (denoted by indices $pq\ldots$), which we  partition into the OMO basis ($mn\ldots$) and the orthogonal complement orbitals, or OCOs ($uv\ldots$). For complete bases, we can write
\begin{align}
    \phi_p(\b{x}) = \sum_q \phi_q(\b{x}_0) U_{qp}(\b{x}), \quad U_{qp}(\b{x}) = \Dbraket{\phi_q(\b{x}_0)}{\phi_p(\b{x})}.
\end{align}
Occupation number states at $\b{x}$ can thus be expressed as
\begin{align}
    \ket{\Phi(\b{x})} = U(\b{x}) \ket{\Phi(\b{x}_0)}, \label{eq:unitary_ON_state}
\end{align}
with
\begin{align}
    U(\b{x}) = \exp(-b(\b{x})), \quad b(\b{x}) = \sum_{pq} b_{pq}(\b{x}) a_p\adj(\b{x}_0) a_q(\b{x}_0),
\end{align}
where $b(\b{x})$ is the anti-Hermitian operator with $b_{pq}(\b{x})$ defined such that $\b{U}(\b{x}) = \exp(-\b{b}(\b{x}))$. The many-body operators can be expanded as
\begin{align}
    a_p\adj(\b{x}) = \sum_q a_q\adj(\b{x}_0) U_{qp}(\b{x}).
\end{align}
To evaluate derivatives with respect to some specific $x$, we expand operators about $\b{x}_0$,
\begin{align}
    a_p\adj(\b{x}) &= a_p\adj + a_p^{(1)\dagger} \Delta x + \frac{1}{2} a_p^{(2)\dagger} (\Delta x)^2 + \ldots \\
    b(\b{x}) &= b^{(1)} \Delta x + \frac{1}{2} b^{(2)} (\Delta x)^2 + \ldots,
\end{align}
where
\begin{align}
    b^{(n)} = \sum_{pq} b^{(n)}_{pq} a_p\adj a_q. 
\end{align}
Here we have let $a_p\adj \equiv a_p\adj(\b{x}_0)$ and suppressed the $\b{x}$-dependence of the derivatives. It will be useful to split operator contributions in the OMO ($\parallel$) and OCO blocks ($\perp$):
\begin{align}
    a_p\adj(\b{x}) = \sum_m a_m\adj(\b{x}_0) U_{mp}(\b{x}) + \sum_u a_u\adj(\b{x}_0) U_{up}(\b{x}) = a_{p\parallel}\adj(\b{x}) + a_{p\perp}\adj(\b{x}).
\end{align}

Let us evaluate
\begin{align}
    f_{IJ} &= \Dbraket{\Phi_I(\b{x}_0)}{\frac{\partial}{\partial x} \Phi_J(\b{x})}\Big\vert_0.
\end{align}
Using Eq.~\eqref{eq:unitary_ON_state}, we get
\begin{align}
    f_{IJ} &= \Tbraket{\Phi_I(\b{x}_0)}{\frac{\partial U}{\partial x}\Big\vert_0}{ \Phi_J(\b{x}_0)} = - \Tbraket{\Phi_I(\b{x}_0)}{b^{(1)}}{\Phi_J(\b{x}_0)}.
\end{align}
To simplify further, we note that $U_{mn} = \Delta_{mn}$ is Hermitian in the natural connection. Since the $U_{uv}$ block can similarly be chosen to be Hermitian, we have\cite{olsen1995orbital}
\begin{align} 
b_{mn} = b_{uv} = 0 \implies b_{mn}^{(k)} = b_{uv}^{(k)} = 0 \label{eq:b_mn_uv}
\end{align}
and so
\begin{align}
    f_{IJ} &= - \sum_{mn} b_{mn}^{(1)} \Tbraket{\Phi_I(\b{x}_0)}{a_m\adj a_n}{\Phi_J(\b{x}_0)} = 0.
\end{align}
In general, $f_{IJ}$ is non-zero with connections other than the natural connection. 

Next, we consider the second derivative
\begin{align}
    g_{IJ} &= \Dbraket{\Phi_I(\b{x}_0)}{\frac{\partial^2}{\partial x^2} \Phi_J(\b{x})}\Big\vert_0 = \Tbraket{\Phi_I(\b{x}_0)}{\frac{\partial^2 U}{\partial x^2}\Big\vert_0}{\Phi_J(\b{x}_0)},
\end{align}
which can be written
\begin{align}
\begin{split} 
    g_{IJ} &= \Tbraket{\Phi_I(\b{x}_0)}{-b^{(2)} + b^{(1)}b^{(1)}}{\Phi_J(\b{x}_0)} = \Tbraket{\Phi_I(\b{x}_0)}{b^{(1)}b^{(1)}}{\Phi_J(\b{x}_0)}.
\end{split}
\end{align}
In the final equality, we have used Eq.~\eqref{eq:b_mn_uv}. Now, notice that since
\begin{align}
    b^{(1)} = \sum_{um} b_{um}^{(1)} a_u\adj a_m + \sum_{mu} b_{mu}^{(1)} a_m\adj a_u = - \sum_m a_{m\perp}^{(1)\dagger} a_m + \sum_m a_m\adj a_{m\perp}^{(1)},
\end{align}
the only non-zero $b^{(1)}b^{(1)}$ contribution is the one that first creates an electron in the complementary space and then destroys it. Thus,
\begin{align}
\begin{split}
    g_{IJ} &= -\sum_{mn} \Tbraket{\Phi_I(\b{x}_0)}{a_m\adj a_{m\perp}^{(1)} a_{n\perp}^{(1)\dagger} a_n}{\Phi_J(\b{x}_0)} \\
    &= -\sum_{mn} \Tbraket{\Phi_I(\b{x}_0)}{a_m\adj [a_{m\perp}^{(1)}, a_{n\perp}^{(1)\dagger}]_+ a_n}{\Phi_J(\b{x}_0)}.
\end{split}
\end{align}
The commutator can be expressed as
\begin{align}
    [a_{m\perp}^{(1)}, a_{n\perp}^{(1)\dagger}]_+ =  \sum_{uv} U^{(1)\ast}_{um} U^{(1)}_{vn} [a_u, a_v\adj]_+ = \sum_u U_{um}^{(1)\ast} U_{un}^{(1)} = \sum_u \Dbraket{\phi_m^{(1)}}{\phi_u(\b{x}_0)}\Dbraket{\phi_u(\b{x}_0)}{\phi_n^{(1)}}. \label{eq:commutator_w_inner_proj}
\end{align}
Moreover, since 
\begin{align}
    \Dbraket{\phi_m(\b{x}_0)}{\phi_n^{(1)}} = U_{mn}^{(1)} = -b_{mn}^{(1)} = 0,
\end{align}
the inner projection in Eq.~\eqref{eq:commutator_w_inner_proj} is equivalent to the identity and so
\begin{align}
    [a_{m\perp}^{(1)}, a_{n\perp}^{(1)\dagger}]_+ = \Dbraket{\phi_m^{(1)}}{\phi_n^{(1)}}.
\end{align}
Hence, we get the final result
\begin{align}
    g_{IJ} = -\sum_{mn} \Tbraket{\Phi_I(\b{x}_0)}{a_m\adj a_n}{\Phi_J(\b{x}_0)} \Dbraket{\phi_m^{(1)}}{\phi_n^{(1)}}.
\end{align}
The formulas for $f_{IJ}$ and $g_{IJ}$ are valid for occupation number states but allow for generalization to general wave functions. We will be concerned with evaluating partial derivatives with respect to $x$ for wave functions of the form
\begin{align}
    \ket{\psi_k(\b{x})} = \sum_I c_{Ik}(\b{x}) \ket{\Phi_I(\b{x})}.
\end{align}
Since the $c_{Ik}$ depend implicitly on $\b{x}$, we have $\partial c_{Ik} / \partial x = 0$. Thus,
\begin{align}
\begin{split} 
    f_{kl} &= \Tbraket{\psi_k(\b{x}_0)}{\frac{\partial}{\partial x}}{\psi_l(\b{x})} \Big\vert_0 \\
    &= \sum_{IJ} c_{Ik}^{\ast}(\b{x}_0)  \Tbraket{\Phi_I(\b{x}_0)}{\frac{\partial}{\partial x}}{\Phi_J(\b{x})} \Big\vert_0 c_{Jl}(\b{x}_0) \\ &= \sum_{IJ} c_{Ik}^{\ast}(\b{x}_0)  f_{IJ} c_{Jl}(\b{x}_0) \\
    &= 0
\end{split}
\end{align}
and
\begin{align}
\begin{split}
    g_{kl} &= \Tbraket{\psi_k(\b{x}_0)}{\frac{\partial^2}{\partial x^2}}{\psi_l(\b{x})} \Big\vert_0 \\
    &= \sum_{IJ} c_{Ik}^{\ast}(\b{x}_0)  g_{IJ} c_{Jl}(\b{x}_0) \\
    &= -\sum_{mn} \Tbraket{\psi_k(\b{x}_0)}{a_m\adj a_n}{\psi_l(\b{x}_0)} \Dbraket{\phi_m^{(1)}}{\phi_n^{(1)}}.
\end{split}
\end{align}

For partial derivatives of the energy, we also have to account for the explicit $\b{x}$-dependence of the Hamiltonian. We express the OMO Hamiltonian as 
\begin{align}
    H = \sum_{pq} h_{pq}(\b{x}) E_{pq}(\b{x}) + \frac{1}{2} \sum_{pqrs} g_{pqrs}(\b{x}) e_{pqrs}(\b{x}),
\end{align}
where both the integrals and the operators depend on $\b{x}$. However, the dependence of the operators can be ignored in energy derivatives because matrix elements of occupation number states are constant:
\begin{align}
    \Dbraket{\Phi_I(\b{x})}{\Phi_J(\b{x})} = \Tbraket{\Phi_I(\b{x}_0)}{U(\b{x})\adj U(\b{x})}{\Phi_J(\b{x}_0)} = \Dbraket{\Phi_I(\b{x}_0)}{\Phi_J(\b{x}_0)}
\end{align}
In particular, elements involving $E_{pq}(\b{x})$ and $e_{pqrs}(\b{x})$ are linear combinations of such  overlaps and therefore give no contributions in energy derivatives\cite{Helgaker1992}. The integrals are related to the UMO basis as
\begin{align}
    h_{pq}(\b{x}) &= \sum_{mn} T_{mp}(\b{x})^\ast h_{mn}^{u}(\b{x}) T_{nq}(\b{x}) \\
    g_{pqrs}(\b{x}) &= \sum_{mnkl} T_{mp}(\b{x})^\ast T_{nq}(\b{x})^\ast g_{mnkl}^{u}(\b{x}) T_{kr}(\b{x}) T_{ls}(\b{x}).
\end{align}
By differentiating $\b{T} \b{W} = \b{W}\adj \b{T}\adj$ and $\b{I} = \b{T}\adj \b{S} \b{T}$, we find that 
\begin{align}
    \b{T}^{(1)} = - \b{W}^{(1)}.
\end{align}
Consequently, the partial derivative of the Hamiltonian can be written
\begin{align}
    H^{(1)} = H^{(1)}_u - \{ W^{(1)}, H \},
\end{align}
where $H^{(1)}_u$ is the derivative of the UMO Hamiltonian and
\begin{align} 
\{ W^{(1)}, H \} = \sum_{pq} j_{pq} E_{pq} + \frac{1}{2} \sum_{pqrs} j_{pqrs} e_{pqrs}, 
\end{align}
where 
\begin{align}
    j_{pq} &= \sum_{m} (W^{(1)}_{pm} h_{mq} + h_{pm} W^{(1)}_{mq}) \\
    j_{pqrs} &= \sum_{m} (W^{(1)}_{pm} g_{mqrs} + W^{(1)}_{qm} g_{pmrs} + g_{pqms} W^{(1)}_{mr} + g_{pqrm} W^{(1)}_{ms}).
\end{align}
The $\b{W}^{(1)}$ matrix, given by
\begin{align}
    W^{(1)}_{pq} = \frac{\partial W_{pq}}{\partial x} \Big\vert_0 = \sum_{\alpha\beta} C_{\alpha p}(\b{x}_0) C_{\beta q}(\b{x}_0) \int \chi_{\alpha}(\b{x}_0) \frac{\partial \chi_\beta}{\partial x} \Big\vert_0 \df \b{r},
\end{align}
is analogous to $S^{(1)}_{pq} = \partial S_{pq} / \partial x \vert_0$ in the symmetric connection $\b{T} = \b{S}^{-1/2}$. 

This concludes our discussion of how the geometry dependence of the many-body operators affects energy derivatives and nonadiabatic coupling elements. We refer the reader to Helgaker and Jørgensen\cite{Helgaker1992} for more details regarding connections and energy derivatives and to Olsen \emph{et al}.\cite{olsen1995orbital} for more on the natural connection. In the next section, we derive expressions for the nonadiabatic elements in coupled cluster theory.

\subsection{Nonadiabatic coupled cluster couplings in a Lagrangian formalism}
To obtain a Lagrangian for the vector coupling, Hohenstein\citep{Hohenstein2016} defined a quantity whose first derivatives are identical to components of the vector coupling. Although Hohenstein used it for  configuration interaction theory, the observation generalizes straightforwardly to coupled cluster theory. The quantity is the partially frozen overlap 
\begin{align}
    \mathscr{O}_{mn}(\b{x}) = \Dbraket{\tilde{\psi}_m(\b{x}_0)}{\psi_n(\b{x})},
\end{align}
in terms of which we have
\begin{align}
    (\b{F}_{mn}^I)_i = \DBraket{\tilde{\psi}_m(\b{x})}{\frac{\df}{\df x_i} \psi_n(\b{x})}\Big\vert_0 = \frac{\df}{\df x_i} \mathscr{O}_{mn}(\b{x})\Big\vert_0, \quad i \in I, \label{eq:hohenstein_vector}
\end{align}
and
\begin{align}
    G_{mn}^I = \DBraket{\tilde{\psi}_m(\b{x})}{\sum_{i\in I} \frac{\df^2}{\df x_i^2} \psi_n(\b{x})}\Big\vert_0 = \sum_{i\in I} \frac{\df^2}{\df x_i^2} \mathscr{O}_{mn}(\b{x}) \Big\vert_0. \label{eq:hohenstein_scalar}
\end{align}
For convenience, we write $i \in I$ to signify that $x_i$ is one of the three coordinates at nucleus $I$ ($x, y,$ or $z$). Clearly, the vector and scalar couplings are derivatives of the partially frozen overlap and may therefore be evaluated using a Lagrangian. Note that the overlap $\mathscr{O}_{mn}(\b{x})$ depends on $\b{x}_0$. We suppress this dependency for notational simplicity.

The overlap is expressed in terms of coupled cluster wave functions, which depend on $\b{x}$ but also on a set of wave function parameters $\b{\lambda}$ (which themselves depend on $\b{x}$). Written out in terms of wave function parameters, the overlap is given by
\begin{align}
    \mathscr{O}_{mn}(\b{x}, \b{\lambda}) = \Tbraket{\tilde{\psi}_m(\b{x}_0)}{\exp(-\kappa)\exp(T)}{\st{R}_n},
\end{align}
where
\begin{align}
    \bra{\tilde{\psi}_m(\b{x}_0)} = \bra{\st{L}_m} \exp(-T) \Big\vert_0
\end{align}
and
\begin{align}
    \kappa = \sum_{p > q} \kappa_{pq} (E_{pq} - E_{qp}) = \sum_{p > q} \kappa_{pq} E_{pq}^-.
\end{align}
The $\kappa$ operator accounts for orbital rotations, meaning changes in the Hartree-Fock orbitals, where, by definition, we have $\kappa(\b{x}_0) = 0$. Following the standard recipe, we add the equations (denoted by $\mathscr{E}_{mn}$) that determine the parameters as constraints with associated Lagrangian multipliers (denoted by $\b{\gamma}$),
\begin{align}
    \mathscr{L}_{mn}(\b{\lambda}, \b{x}, \b{\gamma}) = \mathscr{O}_{mn}(\b{\lambda}, \b{x}) + \b{\gamma}^T \mathscr{E}_{mn}(\b{\lambda}, \b{x}), \quad m \neq n,
\end{align}
where $\b{\lambda}$ and $\b{\gamma}$ are determined for every $\b{x}$ by stationarity:
\begin{align}
    \frac{\partial \mathscr{L}_{mn}}{ \partial \gamma_k}  &= (\mathscr{E}_{mn})_k = 0 \label{eq:stationarity_multipliers} \\
    \frac{\partial \mathscr{L}_{mn}}{ \partial \lambda_k} &= 0. \label{eq:stationarity_parameters}
\end{align}
The derivatives of this Lagrangian are identical to the derivatives of the frozen overlap (since $\mathscr{E}_{mn} = 0$). One advantage of the Lagrangian formalism is that it automatically incorporates the $2n + 1$ and $2n + 2$ rules for $\b{\lambda}$ and $\b{\gamma}$, respectively. 
In particular, 
\begin{align}
    (\b{F}_{mn}^I)_i = \frac{\df \mathscr{L}_{mn}}{\df x_i}\Big\vert_0 = \frac{\partial \mathscr{L}_{mn}}{\partial x_i}\Big\vert_0, \quad i \in I,
\end{align}
where the final equality follows from stationarity, see Eqs.~\eqref{eq:stationarity_multipliers} and \eqref{eq:stationarity_parameters}. Denoting partial derivatives with respect to geometrical coordinates as
\begin{align}
    a^{(i)} = \frac{\partial a}{\partial x_i} \Big\vert_0, \quad a^{(i,j)} = \frac{\partial^2 a}{\partial x_i \partial x_j} \Big\vert_0,
\end{align}
we can write
\begin{align}
    (\b{F}_{mn}^I)_i = \mathscr{L}_{mn}^{(i)}, \quad i \in I.
\end{align}
Furthermore, if we let
\begin{align}
    f_\alpha = \frac{\partial \mathscr{L}_{mn}}{\partial \lambda_\alpha} \Big\vert_{\b{\lambda} = \b{\lambda}_0}, \quad H_{\alpha\beta} =  \frac{\partial^2 \mathscr{L}_{mn}}{\partial \lambda_\alpha \partial \lambda_\beta} \Big\vert_0, \label{eq:gradient_hessian_parameter}
\end{align}
then the scalar coupling can be expressed as (see Appendix A)
\begin{align}
    G_{mn}^I = \sum_{i\in I} \frac{\df^2 \mathscr{L}_{mn}}{\df x_i^2}\Big\vert_0 = \sum_{i \in I} \Bigl( \mathscr{L}_{mn}^{(i,i)} + \sum_{\alpha\beta} \lambda_\alpha^{(i)} H_{\alpha\beta} \lambda_\beta^{(i)} + 2 \sum_\alpha \lambda_\alpha^{(i)} f_\alpha^{(i)} \Bigr). \label{eq:scalar_coupling_Lagrangian}
\end{align}
Clearly, $\b{F}_{mn}^I$ and $G_{mn}^I$ are similar in complexity to the energy gradient and Hessian. However, $G_{mn}^I$ is somewhat simpler than the energy Hessian because the first derivatives of the parameters ($\b{\lambda}^{(i)}$) can be considered one at a time.

To proceed, we must define the Lagrangian $\mathscr{L}_{mn}$ in detail. The conditions $\mathscr{E}_{mn}$ include all equations that must be solved to evaluate the overlap $\mathscr{O}_{mn}$. These are (a) the Hartree-Fock equations, (b) the amplitude equations, and (c) the eigenvalue equations for the right state amplitudes. Written out in full, we have
\begin{align}
\begin{split}
    \mathscr{L}_{mn} &= \mathscr{O}_{mn} + \b{\gamma}^T \mathscr{E}_{mn} \\ 
    &= \mathscr{O}_{mn} + \bar{\b{\kappa}}^T \b{F}_c + \b{\zeta}^T \b{\Omega} +\b{\beta}_n^T(\bar{\bst{H}} - E_n )\bst{R}_n + \bar{E}_n (1 - \Dbraket{\st{L}_n^{0}}{\st{R}_n}), \label{eq:Lagrangian_conditions}
\end{split}
\end{align} 
where we have introduced multipliers associated with the different sets of equations, $\bar{\b{\kappa}}$, $\b{\zeta}$, as well as $\b{\beta}_n$ and $\bar{E}_n$. We have also introduced the Brillouin condition 
\begin{align}
    (\b{F}_c)_{pq} = \Tbraket{\hf}{[E_{pq}^-, H(\kappa)]}{\hf}, \quad p > q,
\end{align}
where
\begin{align}
    H(\kappa) = \exp(\kappa) H \exp(-\kappa).
\end{align}
Furthermore, the similarity transformed Hamiltonian in $\b{\Omega}$ and $\bar{\bst{H}}$ is given by 
\begin{align}
    \bar{H} = \bar{H}(\kappa) = \exp(-T) H(\kappa) \exp(T)
\end{align}
and the $n$th electronic energy defined as
\begin{align}
    E_n = \Tbraket{\st{L}_n^{(0)}}{\bar{H}}{\st{R}_n}.
\end{align}

With $\mathscr{L}_{mn}$ defined, we can now consider the equations for the zeroth order multipliers. These are determined from the zeroth order terms of the $\b{\lambda}$ stationarity, Eq.~\eqref{eq:stationarity_parameters}. To keep our notation simple, we will denote the zeroth order terms as $\b{\gamma}^{(0)} \equiv \b{\gamma}$ and $\b{\lambda}^{(0)} \equiv \b{\lambda}$, where it should be understood from context when these are $\b{\gamma}$ and $\b{\lambda}$ evaluated at $\b{x}_0$. Differentiation with respect to the state parameters gives
\begin{align}
    \frac{\partial \mathscr{L}_{mn}}{\partial \bst{R}_n}\Big\vert_0 &= 0 = \bst{L}_m^T + \b{\beta}_n^T (\bar{\bst{H}} - E_n \b{I}) - \bigl(\bar{E}_n + E_n \, \b{\beta}_n^T \b{R}_n \bigr)\bst{L}_n^T = 0. \label{eq:stationarity_right_amplitudes}
\end{align}
To solve this equation, we note that if we let
\begin{align}
    \bar{E}_n = - E_n \, \b{\beta}_n^T \b{R}_n,
\end{align}
the equation for $\b{\beta}_n$ becomes
\begin{align}
    \b{\beta}_n^T (\bar{\bst{H}} - E_n \b{I}) &= -\bst{L}_m^T. 
\end{align}
Thus, we have
\begin{align}
    \b{\beta}_n &= (E_n - E_m)^{-1} \bst{L}_m.
\end{align}

Next we consider stationarity with respect to $\b{t}$. This can be expressed as 
\begin{align}
    \frac{\partial \mathscr{L}_{mn}}{\partial \b{t}}\Big\vert_0 = 0 = {^t}\b{\eta} + \b{\zeta}^T \b{A}, \label{eq:stationarity_cluster}
\end{align}
where
\begin{align}
    {^{t}}\eta_\mu = \Tbraket{\bar{\st{L}}_m}{\tau_\mu}{\bar{\st{R}}_n} + 
    \Tbraket{\bar{\beta}_n}{[H,\tau_\mu]}{\bar{\st{R}}_n}
\end{align}
and where we have introduced the notation
\begin{align}
    \ket{\bar{X}} &= \exp(T)\ket{X} \\
    \bra{\bar{X}} &= \bra{X} \exp(-T).
\end{align}

Finally, we have stationarity with respect to $\b{\kappa}$, which can be written
\begin{align}
    \frac{\partial \mathscr{L}_{mn}}{\partial \b{\kappa}}\Big\vert_0 = 0 = {^\kappa}\b{\eta} +  \bar{\b{\kappa}}^T \b{A}^\mathrm{HF},\label{eq:stationarity_kappa}
\end{align}
where
\begin{align}
\begin{split} 
    {^{\kappa}}\eta_{rs} = &-\Tbraket{\bar{\st{L}}_m}{ E_{rs}^- }{\bar{\st{R}}_n} + \Tbraket{\bar{\zeta}}{[E_{rs}^-,H]}{\cc} + \Tbraket{\bar{\beta}_n}{ [E_{rs}^-, H] }{\bar{\st{R}}_n}
\end{split}
\end{align}
and
\begin{align}
    A_{pq,rs}^\mathrm{HF} &= \Tbraket{\hf}{[E_{pq}^-,[E_{rs}^-,H]]}{\hf}.
\end{align}

With the zeroth order multipliers determined, we can derive the expression for the vector coupling. By partially differentiating $\mathscr{L}_{mn}$, we find that
\begin{align}
\begin{split}
    (\b{F}_{mn}^I)_i = \mathscr{L}_{mn}^{(i)} &= (E_n - E_m)^{-1} \Tbraket{\st{L}_m}{\bar{H}^{(i)}}{\st{R}_n} \\
    &+ \Tbraket{\zeta}{\bar{H}^{(i)}}{\hf} + \Tbraket{\hf}{[\bar{\kappa},H^{(i)}]}{\hf}, \quad i \in I,
\end{split} \label{eq:derivative_coupling_cc}
\end{align}
where
\begin{align}
    \bar{H}^{(i)} = \exp(-T) H^{(i)} \exp(T)
\end{align}
and where quantities at $\b{x}_0$ are denoted as $y^{(0)} \equiv y$ (e.g., we denote $T^{(0)}$ as $T$). 

The  vector coupling given in Eq.~\eqref{eq:derivative_coupling_cc} has also been identified by other authors. It was derived by Christiansen,\cite{Christiansen1999nonadiabatic} who assumed biorthonormality and used $Z$-vector substitution\cite{handy1984evaluation} on the expression for the vector coupling. Tajti and Szalay\cite{Tajti2009} identified the same expression indirectly via $Z$-vector substitution on derivatives of Hamiltonian transition elements. However, they also argued\cite{Tajti2009} that the coupling should not be given by Eq.~\eqref{eq:derivative_coupling_cc} but rather be averaged and expressed with normalized states. As we have shown, Eq.~\eqref{eq:derivative_coupling_cc} is a valid choice due to norm invariance and represents the vector coupling in the right nuclear Schrödinger equations. 
For the left Schrödinger equations, we can make use of the identity
\begin{align}
    \Dbraket{\tilde{\psi}_m}{\psi_n} = \delta_{mn} \implies \b{F}_{mn}^I = - \tilde{\b{F}}_{mn}^I. \label{eq:vector_to_vector_tilde}
\end{align} 
Before moving on to the scalar coupling, we note that although the 
$Z$-vector substitution method is equivalent to the Lagrangian technique, the latter method gives, in our opinion, an especially elegant way of deriving the  coupling elements.

For the scalar coupling, we must determine the first derivatives of the parameters. Equations for these are obtained as the first order terms of the multiplier stationarity conditions. In the case of $\b{t}$, we have
\begin{align}
    \frac{\df \b{\Omega}}{\df x_i} \Big\vert_0 = 0 = {^t}\b{\xi}^{(i)} + \b{A} \b{t}^{(i)},
\end{align}
where
\begin{align}
    {^t}\xi_\mu^{(i)} = \Tbraket{\mu}{\bar{H}^{(i)}}{\hf} + \Tbraket{\mu}{\exp(-T)[\kappa^{(i)},H]\exp(T)}{\hf}.
\end{align}
In the case of $\b{\kappa}$, we similarly have
\begin{align}
    \frac{\df \b{F}_c}{\df x_i}\Big\vert_0 = 0 = {^\kappa}\b{\xi}^{(i)} + \b{A}^\mathrm{HF} \b{\kappa}^{(i)},
\end{align}
where
\begin{align}
    {^\kappa}\xi^{(i)}_{rs} = \Tbraket{\hf}{[E_{rs}^-, H^{(i)}]}{\hf}.
\end{align}
The binormality condition implies
\begin{align}
    \frac{\df}{\df x_i} ( 1 - \Dbraket{\st{L}_n^{0}}{\st{R}_n} ) \Big\vert_0 = 0 = - \bst{L}_n^T \bst{R}_n^{(i)}, \label{eq:binormality_1st_response}
\end{align}
while the eigenvalue condition implies
\begin{align}
    \frac{\df}{\df x_i} \Bigl( (\bar{\bst{H}} - E_n )\bst{R}_n \Bigr) \Big\vert_0 &= 0 = {^{\st{R}_n}}\b{\xi}^{(i)} + (\bar{\bst{H}} - E_n)\bst{R}_n^{(i)} \label{eq:response_R}
\end{align}
Here we have defined
\begin{align}
    {^{\st{R}_n}}\b{\xi}^{(i)} &= \Bigl( \bst{Y}^{(i)} - \bst{L}_n^T \bst{Y}^{(i)} \bst{R}_n \Bigr) \bst{R}_n 
\end{align}
where
\begin{align}
    \st{Y}^{(i)}_{\mu\nu} = \Tbraket{\mu}{\bar{H}^{(i)}}{\nu} + \Tbraket{\mu}{[\bar{H},T^{(i)}]}{\nu} + \Tbraket{\mu}{[\kappa^{(i)},\bar{H}]}{\nu}, \quad \mu,\nu \geq 0.
\end{align}

With the derivatives of the parameters determined, let us next consider $f_\alpha^{(i)}$ and $H_{\alpha\beta}$, see Eq.~\eqref{eq:gradient_hessian_parameter}. Recall that the $\alpha$ and $\beta$ indices refer to the parameters $\lambda_{\alpha}$ and $\lambda_{\beta}$. The gradient $\b{f}$ is given by the zeroth order equations for the multipliers, that is, Eqs.~\eqref{eq:stationarity_right_amplitudes}, \eqref{eq:stationarity_cluster}, and \eqref{eq:stationarity_kappa}, with $\b{\lambda} = \b{\lambda}_0$ but allowing for $\b{x} \neq \b{x}_0$. 
Partially differentiating these terms with respect to $x_i$ gives $\b{f}
^{(i)}$. The blocks of the $\sum_{\alpha} f_\alpha^{(i)} \lambda_\alpha^{(i)}$ contributions to $G_{mn}^I$ may be written
\begin{align}
    {^{\st{R}_n}}f^{(i)}_\mu (\bst{R}_n^{(i)})_\mu = \frac{\partial^2 \mathscr{L}_{mn}}{\partial x_i \partial \st{R}_\mu^n} \Big\vert_0  (\bst{R}_n^{(i)})_\mu &=
    \Tbraket{\bar{\beta}_n}{H^{(i)} - E_n^{(i)}}{\bar{\st{R}}_n^{(i)}}
\end{align}
and
\begin{align}
    {^t}f_\mu^{(i)} t_\mu^{(i)} = \frac{\partial^2 \mathscr{L}_{mn}}{\partial x_i \partial t_\mu} \Big\vert_0  t_\mu^{(i)} &= \Tbraket{\bar{\zeta}}{[H^{(i)}, T^{(i)}]}{\cc} + \Tbraket{\bar{\beta}_n}{[H^{(i)}, T^{(i)}]}{\bar{\st{R}}_n}
    \\
\begin{split} 
    {^\kappa}f_{pq}^{(i)} \kappa_{pq}^{(i)} = \frac{\partial^2 \mathscr{L}_{mn}}{\partial x_i \partial \kappa_{pq}} \Big\vert_0  \kappa_{pq}^{(i)} &= \Tbraket{\bar{\zeta}}{[\kappa^{(i)},H^{(i)}]}{\cc} + 
    \Tbraket{\bar{\beta}_n}{[\kappa^{(i)},H^{(i)}]}{\bar{\st{R}}_n} \\
    &+ \Tbraket{\hf}{[\bar{\kappa},[\kappa^{(i)}, H^{(i)}]]}{\hf}
\end{split}
\end{align}
where repeated indices implies summation. For contributions to $G_{mn}^I$ involving the parameter Hessian $H_{\alpha\beta} = \partial^2 \mathscr{L}_{mn} / \partial \lambda_\alpha \partial \lambda_\beta \vert_0$, we have, for terms involving $\b{t}$ and $\b{\kappa}$,
\begin{align}
\begin{split}
     t_\mu^{(i)} H_{\mu\nu}^{tt} t_\nu^{(i)} &= t_\mu^{(i)} \frac{\partial^2 \mathscr{L}_{mn}}{\partial t_\mu \partial t_\nu} \Big\vert_0 t_\nu^{(i)} \\ 
     &= \Tbraket{\bar{\st{L}}_m}{(T^{(i)})^2}{\bar{\st{R}}_n} \\
     &+ \Tbraket{\bar{\zeta}}{[[H, T^{(i)}],T^{(i)}]}{\cc} \\
    &+ 
    \Tbraket{\bar{\beta}_n}{[[H, T^{(i)}], T^{(i)}]}{\bar{\st{R}}_n}
\end{split}
\end{align}
as well as
\begin{align} 
\begin{split}
    \kappa_{rs}^{(i)} H_{rs\nu}^{\kappa t} t_\nu^{(i)} &= \kappa_{rs}^{(i)}\frac{\partial^2 \mathscr{L}_{mn}}{\partial \kappa_{rs} \partial t_\nu} \Big\vert_0 t_\nu^{(i)} \\ 
    &= -\Tbraket{\bar{\st{L}}_m}{\kappa^{(i)} T^{(i)}}{\bar{\st{R}}_n} \\
    &+ \Tbraket{\bar{\zeta}}{[[\kappa^{(i)},H],T^{(i)}]}{\cc} \\
    &+  \Tbraket{\bar{\beta}_n}{[[\kappa^{(i)},H],T^{(i)}]}{\bar{\st{R}}_n}
\end{split}
\end{align}
and
\begin{align} 
\begin{split}
    \kappa_{pq}^{(i)} H_{pqrs}^{\kappa \kappa} \kappa_{rs}^{(i)} &= \kappa_{pq}^{(i)}\frac{\partial^2 \mathscr{L}_{mn}}{\partial \kappa_{pq} \partial \kappa_{rs}} \Big\vert_0 \kappa_{rs}^{(i)} \\
    &= \Tbraket{\bar{\st{L}}_m}{(\kappa^{(i)})^2}{\bar{\st{R}}_n} \\
    &+ \Tbraket{\bar{\zeta}}{[\kappa^{(i)},[\kappa^{(i)},H]]}{\cc} \\
    &+ \Tbraket{\bar{\beta}_n}{[\kappa^{(i)},[\kappa^{(i)},H]]}{\bar{\st{R}}_n} \\
    &+ \Tbraket{\hf}{[\bar{\kappa},[\kappa^{(i)},[\kappa^{(i)}, H]]]}{\hf}.
\end{split} 
\end{align}

Next we have terms involving right state and the cluster amplitudes and orbital rotations:
\begin{align}
    \begin{split}
        t_\mu^{(i)} H_{\mu \nu}^{t \st{R}_n}  (\bst{R}_n^{(i)})_\nu &= t_{\mu}^{(i)} \frac{\partial^2 \mathscr{L}_{mn}}{\partial t_{\mu} \partial (\bst{R}_n)_\nu} \Big\vert_0 (\bst{R}_n^{(i)})_\nu \\
        &=  \Tbraket{\bar{\st{L}}_m}{T^{(i)}}{\bar{\st{R}}_n^{(i)}} \\ &+\Tbraket{\bar{\beta}_n}{[H, T^{(i)}]}{\bar{\st{R}}_n^{(i)}}  \\
        &- \Dbraket{\beta_n}{\st{R}_n^{(i)}}\Tbraket{\bar{\st{L}}_n}{[H, T^{(i)}]}{\bar{\st{R}}_n} 
    \end{split} \\
    \begin{split}
        \kappa_{pq}^{(i)} H_{pq \nu}^{\kappa \st{R}_n}  (\bst{R}_n^{(i)})_\nu &= \kappa_{pq}^{(i)} \frac{\partial^2 \mathscr{L}_{mn}}{\partial \kappa_{pq} \partial (\bst{R}_n)_\nu} \Big\vert_0 (\bst{R}_n^{(i)})_\nu \\ 
        &= -  \Tbraket{\bar{\st{L}}_m}{\kappa^{(i)}}{\bar{\st{R}}_n^{(i)}} \\ &+ \Tbraket{\bar{\beta}_n}{[\kappa^{(i)}, H]}{\bar{\st{R}}_n^{(i)}}  \\
        &- \Dbraket{\beta_n}{\st{R}_n^{(i)}}\Tbraket{\bar{\st{L}}_n}{[\kappa^{(i)},H]}{\bar{\st{R}}_n}. 
    \end{split}
\end{align}
Finally, we have the partial derivative of the Lagrangian, which can be written
\begin{align}
\begin{split}
    \mathscr{L}^{(i,i)}_{mn} &= - \sum_{pq} \Tbraket{\bar{\st{L}}_m}{E_{pq}}{\bar{\st{R}}_n} \Dbraket{\phi_p^{(i)}}{\phi_q^{(i)}} +  \Tbraket{\bar{\zeta}}{H^{(i,i)}}{\cc} \\
    &+ \Tbraket{\hf}{[\bar{\kappa}, H^{(i,i)}]}{\hf} +  \Tbraket{\bar{\beta}_n}{H^{(i,i)}}{\bar{\st{R}}_n},
\end{split}
\end{align}

Written in compact notation, the scalar coupling may be evaluated as
\begin{align}
    \begin{split}
        G_{mn}^I &= \Dbraket{\bar{\st{L}}_m}{\bar{R}_n^{(i,i)}} +\Tbraket{\bar{\zeta}}{K^{(i,i)}}{\cc} 
        + \Tbraket{\bar{\beta}_n}{K^{(i,i)}}{\bar{\st{R}}_n} \\
        &+ \Tbraket{\hf}{[\bar{\kappa}, J^{(i,i)}]}{\hf}
+ \Tbraket{\bar{\beta}_n}{L^{(i)} - \braket{L^{(i)}}_n}{\bar{\st{R}}_n^{(i)}}
    \end{split} \label{eq:scalar_coupling}
\end{align}
where we have let
\begin{align}
    K^{(i,i)} &= J^{(i,i)} + 2 [J^{(i)},T^{(i)}] + [[H, T^{(i)}],T^{(i)}] \\
    J^{(i)} &= H^{(i)} + [\kappa^{(i)}, H] \\
    L^{(i)} &= 2 (H^{(i)} + [\kappa^{(i)}, H] + [H, T^{(i)}]) \\
    \braket{L^{(i)}}_n &= \Tbraket{\bar{\st{L}}_n}{L^{(i)}}{\bar{\st{R}}_n} \\
    J^{(i,i)} &= H^{(i,i)} + 2 [\kappa^{(i)}, H^{(i)}] + [\kappa^{(i)},[\kappa^{(i)}, H]] 
\end{align}
as well as
\begin{align}
\begin{split} 
    \Dbraket{\bar{\st{L}}_m}{\bar{R}_n^{(i,i)}} &= \Tbraket{\bar{\st{L}}_m}{(\kappa^{(i)})^2 - 2\kappa^{(i)} T^{(i)} + (T^{(i)})^2}{\bar{\st{R}_n}} -  2\Tbraket{\bar{\st{L}}_m}{\kappa^{(i)}}{\bar{\st{R}}_n^{(i)}} \\
    &+ 2\Tbraket{\bar{\st{L}}_m}{T^{(i)}}{\bar{\st{R}}_n^{(i)}} - \sum_{pq} \Tbraket{\bar{\st{L}}_m}{E_{pq}}{\bar{\st{R}}_n} \Dbraket{\phi_p^{(i)}}{\phi_q^{(i)}}.
\end{split} \label{eq:derivative_terms}
\end{align}

Throughout the derivations above, we have considered the off-diagonal coupling elements ($m \neq n$). The diagonal terms can be derived from the Lagrangian 
\begin{align}
\begin{split}
    \mathscr{L}_{nn} &= \mathscr{O}_{nn} + \b{\gamma}^T \mathscr{E}_{nn} \\ 
    &= \mathscr{O}_{nn} + \bar{\b{\kappa}}^T \b{F}_c + \b{\zeta}^T \b{\Omega} +\b{\beta}_n^T(\bar{\bst{H}} - E_n )\bst{R}_n + \bar{E}_n (1 - \Dbraket{\st{L}_n^{0}}{\st{R}_n}), \label{eq:Lagrangian_conditions_diagonal}
\end{split}
\end{align} 
which gives the slightly different $\st{R}_n$ stationarity condition
\begin{align}
    0 = (1 + \bar{E}_n - E_n \b{\beta}_n^T \bst{R}_n) \bst{L}_n^T + \b{\beta}_n^T(\bar{\bst{H}} - E_n ).
\end{align}
Here we again select $\bar{E}_n$ to make the first term vanish, giving
\begin{align}
    \b{\beta}_n^T = \bst{L}_n^T.
\end{align}
Other than this change, the derivation of the scalar coupling is virtually unchanged. Terms involving differentiation of $\mathscr{O}_{nn}$ has the left state $\bra{\st{L}_n}$ in the bra instead of $\bra{\st{L}_m}$ (e.g., in the stationarity conditions for the zeroth order multipliers). In particular, the expression in Eq.~\eqref{eq:scalar_coupling} is valid with $m = n$. 

Unlike for the vector coupling, there is no convenient relationship between $G_{mn}^I$ and $\tilde{G}_{mn}^I$. To derive the latter quantity, we may consider the Lagrangian
\begin{align}
\begin{split}
    \mathscr{L}_{mn} &= \mathscr{O}_{mn} + \b{\gamma}^T \mathscr{E}_{mn} \\ 
    &= \mathscr{O}_{mn} + \bar{\b{\kappa}}^T \b{F}_c + \b{\zeta}^T \b{\Omega} +\bst{L}_m^T(\bar{\bst{H}} - E_m )\b{\beta}_m + \bar{E}_m (1 - \Dbraket{\st{L}_m}{\st{R}_m^{0}}), \label{eq:Lagrangian_conditions_tilde}
\end{split}
\end{align} 
where
\begin{align}
    \mathscr{O}_{mn} &= \Tbraket{\st{L}_m}{\exp(-T)\exp(\kappa)}{\psi_n(\b{x}_0)} \\
    E_m &= \bst{L}_m^T \bar{\bst{H}} \bst{R}_m^0.
\end{align}
The $\st{L}_m$ stationarity then gives
\begin{align}
    0 = \bst{R}_n + (\bar{\bst{H}} - E_m )\b{\beta}_m + (\bar{E}_m - E_m \bst{L}_m^T \b{\beta}_m) \bst{R}_m,
\end{align}
from which we again have $\bar{E}_m = E_m \bst{L}_m^T \b{\beta}_m$ and thus
\begin{align}
    \b{\beta}_m = -(E_n - E_m)^{-1} \bst{R}_n.
\end{align}
The equations for the zeroth order multipliers are derived as before, with the result that the multipliers change their sign, thus giving the result in Eq.~\eqref{eq:vector_to_vector_tilde} for the vector coupling. For the derivative of the parameters, we have the same equations for $\b{t}^{(i)}$ and $\b{\kappa}^{(i)}$. For the derivative of $\st{L}_m$, we must solve the equation
\begin{align}
   \frac{\df}{\df x_i} \Bigl( \bst{L}_m^T(\bar{\bst{H}} - E_m ) \Bigr)\Big\vert_0 = 0 = {^{\st{L}_m}}\b{\xi}^{(i)T} + \bst{L}_m^{(i)T}(\bar{\bst{H}} - E_m),
\end{align}
which is analogous to Eq.~\eqref{eq:response_R}. In contributions involving $\Tbraket{\beta_n}{\ldots}{\st{R}_n}$ in $G_{mn}^I$, we obtain similar expressions involving $\Tbraket{\st{L}_m}{\ldots}{\beta_m}$ in the case of $\tilde{G}_{mn}^I$. The end-result is
\begin{align}
    \begin{split}
        \tilde{G}_{mn}^I &= \Dbraket{\bar{\st{L}}_m^{(i,i)}}{\bar{R}_n} +\Tbraket{\bar{\zeta}}{K^{(i,i)}}{\cc} 
        + \Tbraket{\bar{\st{L}}_m}{K^{(i,i)}}{\bar{\beta}_m} \\
        &+ \Tbraket{\hf}{[\bar{\kappa}, J^{(i,i)}]}{\hf}
+ \Tbraket{\bar{\st{L}}_m^{(i)}}{L^{(i)} - \braket{L^{(i)}}_m}{\bar{\beta}_m},
    \end{split} \label{eq:scalar_coupling_tilde}
\end{align}
with
\begin{align}
    \begin{split}
        \Dbraket{\bar{\st{L}}_m^{(i,i)}}{\bar{R}_n} &= \Tbraket{\bar{\st{L}}_m}{(\kappa^{(i)})^2 - 2  T^{(i)} \kappa^{(i)} + (T^{(i)})^2}{\bar{\st{R}_n}} +  2\Tbraket{\bar{\st{L}}_m^{(i)}}{\kappa^{(i)}}{\bar{\st{R}}_n} \\
    &- 2\Tbraket{\bar{\st{L}}_m^{(i)}}{T^{(i)}}{\bar{\st{R}}_n} - \sum_{pq} \Tbraket{\bar{\st{L}}_m}{E_{pq}}{\bar{\st{R}}_n} \Dbraket{\phi_p^{(i)}}{\phi_q^{(i)}}.
    \end{split}
\end{align}
Finally, $\tilde{G}_{nn}^I$ is obtained in a similar manner to $G_{nn}^I$, see Eq.~\eqref{eq:Lagrangian_conditions_diagonal} and the surrounding text.

This concludes our derivation of the  coupled cluster scalar coupling. To the best of our knowledge, equations for this coupling have not been presented in the literature before (with $m \neq n$). Diagonal terms were also considered by Gauss \emph{et al}.\cite{Gauss2006} from a different starting point.
The scalar coupling is often omitted in dynamics simulations, but  its potential influence on nonadiabatic dynamics has been highlighted in recent years (see Curchod and Martínez\cite{Curchod2018} and references therein). 

\section{Concluding remarks}
The norm of the electronic states changes the value of nonadiabatic coupling elements but does not change the molecular wave function. The biorthonormal formula assumed by Christiansen\cite{Christiansen1999nonadiabatic} is therefore a valid choice for coupled cluster dynamics with the appropriate nuclear Schrödinger equations. More generally, we have shown that the wave function is invariant under invertible transformations of the electronic basis.
Of course, the biorthonormal couplings are not directly comparable to the coupling elements of an Hermitian method with normalized states, such as CI or full-CI. However, this reflects the basis-dependence of the couplings and not the validity of the biorthonormal formalism.

We therefore derive a set of nuclear Schrödinger equations assuming  biorthonormal projection onto the electronic basis. Combined with expressions derived for the vector and scalar couplings, these nuclear Schrödinger equations serve as a starting point for the application of nonadiabatic dynamics methods to coupled cluster theory. 

Our derivations have been restricted to standard coupled cluster theory. However, the Lagrangian formalism is easily extended to similarity constrained coupled cluster methods,\cite{kjoenstad2017resolving,kjoenstad2019invariant} which are suited to describe relaxation through a conical intersection between excited states. The application to ground state intersections is less straightforward, but may be accessible with approaches that use a different reference than the closed-shell Hartree-Fock state.\cite{Lefrancois2017}

\begin{acknowledgement}
We thank Todd J.~Martínez for enlightening discussions. We also acknowledge funding from the Marie Sk{\l}odowska-Curie European Training Network ``COSINE - COmputational Spectroscopy In Natural sciences and Engineering'', Grant Agreement No. 765739, and the Research Council of Norway through FRINATEK projects 263110 and 275506.



\end{acknowledgement}


\section{Appendix A: Lagrangian derivatives}
Here we derive first and second derivatives of the generic Lagrangian
\begin{align}
    \mathscr{L}(\b{\lambda}, \b{x}, \b{\gamma}) = \mathscr{O}(\b{\lambda},\b{x}) + \b{\gamma}^T \mathscr{E}(\b{\lambda}, \b{x})
\end{align}
with respect to $\b{x}$. The parameters and multipliers both depend on $\b{x}$ since they are determined, for a given $\b{x}$, from the stationarity conditions
\begin{align}
    \frac{\partial \mathscr{L}}{\partial \lambda_k} = 0, \quad \frac{\partial \mathscr{L}}{\partial \gamma_k} = 0.
\end{align}
Using Einstein notation, we can write the Taylor expansion of $\mathscr{L}$ about some $\b{x}_0$ as
\begin{align}
\begin{split}
    \mathscr{L}(\b{\lambda}, \b{x}, \b{\gamma}) &= \mathscr{L}_0 
    + \frac{\partial \mathscr{L}}{\partial x_k}\Big\vert_0 \Delta x_k + \frac{1}{2} \Delta x_k \frac{\partial^2 \mathscr{L}}{\partial x_k \partial x_l}\Big\vert_0 \Delta x_l + \frac{1}{2} \Delta \lambda_k \frac{\partial^2 \mathscr{L}}{\partial \lambda_k \partial \lambda_l}\Big\vert_0 \Delta \lambda_l \\
    & + \Delta \lambda_k \frac{\partial^2 \mathscr{L}}{\partial \lambda_k \partial x_l}\Big\vert_0 \Delta x_l + 
    \Delta \gamma_k \frac{\partial^2 \mathscr{L}}{\partial \gamma_k \partial x_l}\Big\vert_0 \Delta x_l + \Delta \lambda_k \frac{\partial^2 \mathscr{L}}{\partial \lambda_k \partial \gamma_l}\Big\vert_0 \Delta \gamma_l + \ldots,
\end{split} \label{eq:Taylor_Lagrangian}
\end{align}
where we have ignored terms of order three or higher in $\Delta \b{x}$. These terms do not contribute to the first and second derivatives and are therefore not relevant to the analysis given here.

In the first derivative, only the partial derivative survives,
\begin{align}
    \frac{\df \mathscr{L}}{\df x_i}\Big\vert_0 = \frac{\partial \mathscr{L}}{\partial x_i}\Big\vert_0.
\end{align}
This is due to the stationarity conditions, since they ensure that there are no linear terms in $\Delta \b{\lambda}$ and $\Delta \b{\gamma}$ in the Taylor expansion in Eq.~\eqref{eq:Taylor_Lagrangian}. In the second derivative, it is convenient to introduce notation for derivatives with respect to particular components of $\b{x}$. We let
\begin{align}
    a^{(i)} &= \frac{\partial a}{\partial x_i} \Big\vert_0 \\
    a^{(i,j)} &= \frac{\partial^2 a}{\partial x_i \partial x_j} \Big\vert_0.
\end{align}
Then we can write
\begin{align}
    \frac{\df \mathscr{L}}{\df x_i}\Big\vert_0 = \mathscr{L}^{(i)}.
\end{align}
and
\begin{align}
\begin{split} 
    \frac{\df^2 \mathscr{L}}{\df x_i \df x_j} \Big\vert_0 &= \mathscr{L}^{(i,j)} + \lambda_k^{(i)} \frac{\partial^2 \mathscr{L}}{\partial \lambda_k \partial \lambda_l}\Big\vert_0 \lambda_l^{(j)} + \lambda_k^{(j)} \frac{\partial^2 \mathscr{L}}{\partial \lambda_k \partial x_i}\Big\vert_0 + \lambda_k^{(i)} \frac{\partial^2 \mathscr{L}}{\partial \lambda_k \partial x_j}\Big\vert_0 \\
    &+ \gamma_k^{(j)} \frac{\partial^2 \mathscr{L}}{\partial \gamma_k \partial x_i}\Big\vert_0 + \gamma_k^{(i)} \frac{\partial^2 \mathscr{L}}{\partial \gamma_k \partial x_j}\Big\vert_0 + 2 \lambda_k^{(i)} \frac{\partial^2 \mathscr{L}}{\partial \lambda_k \partial \gamma_l}\Big\vert_0 \gamma_l^{(j)}.
\end{split}
\end{align}
Now, 
\begin{align}
    \gamma_l^{(j)}\Bigl( \frac{\partial^2 \mathscr{L}}{ \partial x_i \partial \gamma_l}\Big\vert_0 + \lambda_k^{(i)} \frac{\partial^2 \mathscr{L}}{\partial \lambda_k \partial \gamma_l}\Big\vert_0 \Bigr) = \gamma_l^{(j)} \frac{\df \mathscr{E}_l}{\df x_i}\Big\vert_0 = 0,
\end{align}
by stationarity, so that
\begin{align}
    \frac{\df^2 \mathscr{L}}{\df x_i \df x_j} \Big\vert_0 &= \mathscr{L}^{(i,j)} + \lambda_k^{(i)} \frac{\partial^2 \mathscr{L}}{\partial \lambda_k \partial \lambda_l}\Big\vert_0 \lambda_l^{(j)} + \lambda_k^{(j)} \frac{\partial^2 \mathscr{L}}{\partial \lambda_k \partial x_i}\Big\vert_0 + \lambda_k^{(i)} \frac{\partial^2 \mathscr{L}}{\partial \lambda_k \partial x_j}\Big\vert_0.
\end{align}
To simplify the notation further, we define derivatives with respect to the parameters:
\begin{align}
    f_k = \frac{\partial \mathscr{L}}{\partial \lambda_k}, \quad H_{kl} = \frac{\partial^2 \mathscr{L}}{\partial \lambda_k \partial \lambda_l}\Big\vert_0.
\end{align}
Thus, we get the final expression for the second derivatives:
\begin{align}
    \frac{\df^2 \mathscr{L}}{\df x_i \df x_j} \Big\vert_0 = \mathscr{L}^{(i,j)} + \lambda_k^{(i)} H_{kl} \lambda_l^{(j)} + \lambda_k^{(j)} f_k^{(i)} + \lambda_k^{(i)} f_k^{(j)}.
\end{align}

\bibliography{paper}

\clearpage

\centering
\includegraphics{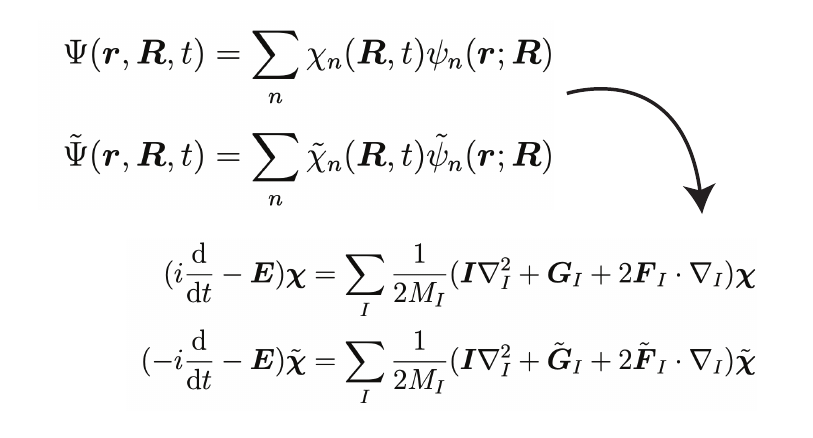}

\end{document}